\documentclass{aastex}
\renewcommand{\cite}{\citealp}

\usepackage{graphics,psfig}
\usepackage{emulateapj5,apjfonts}

\newcommand{\realfigure}[2]{ 
             \hbox{~}
             \centerline{\includegraphics[width=3.5in]{#1}}
             \figcaption{#2}
             \vspace{0.05in}\centerline{}}

\newcommand{\citedraft}[1]{{#1}}

\newcommand{\leoi}{Leo\,{\sc i}}

\newcommand{\leoii}{Leo\,{\sc ii}}
\newcommand{\cmd}{{\sc cmd}}
\newcommand{\wfpred}{{\sc wfpred}}
\newcommand{\mscred}{{\sc mscred}}
\newcommand{\gratis}{{\sc gratis}}
\newcommand{\varfind}{{\sc varfind}}

\newcommand{\allframe}{{\sc allframe}}
\newcommand{\pp}{^\prime}

\submitted{}

\begin{document}


\title{RR Lyrae variables in the dwarf  spheroidal galaxy Leo\,{\sc i}
\altaffilmark{1}}

\author{Enrico V. Held}
\affil{Osservatorio Astronomico di Padova, 
Vicolo dell'Osservatorio 5, 
I-35122 Padova, Italy}
\email{held@pd.astro.it}

\author{Gisella Clementini}
\affil{Osservatorio Astronomico di Bologna, 
Via Ranzani 1, 
I-40127 Bologna, Italy}
\email{gisella@bo.astro.it}

\author{Luca Rizzi\altaffilmark{2}}
\affil{Osservatorio Astronomico di Padova, 
Vicolo dell'Osservatorio 5, I-35122 Padova, Italy}
\email{rizzi@pd.astro.it}

\author{Yazan Momany}
\affil{Dipartimento di Astronomia, Universit\`a di Padova, 
Vicolo dell'Osservatorio 2, I-35122 Padova, Italy}
\email{momany@pd.astro.it}

\author{Ivo Saviane} 
\affil{European Southern Observatory, 
Casilla 19001, Santiago 19, Chile}
\email{isaviane@eso.org}

\and

\author{Luca Di Fabrizio}
\affil{Centro Galileo Galilei \& Telescopio Nazionale Galileo, 
PO Box 565, 38700 S.Cruz de La Palma, Spain}
\email{difabrizio@tng.iac.es}

\altaffiltext{1}{Based on data collected at E.S.O. La Silla, Chile, 
Prop. No. 65.N-0530}
\altaffiltext{2}{also Dipartimento di Astronomia, Universit\`a di Padova}


\begin{abstract}
We report the discovery of a significant population of RR Lyrae
variables in the dwarf spheroidal galaxy \leoi. Based on 40 $V$ and 22
$B$ images of the galaxy taken using the ESO Wide Field Imager we have
identified so far 74 candidate RR Lyrae's in two CCD's hosting the
main body of the galaxy. Full coverage of the light variations and
pulsation periods have been obtained for 54 of them, 47 of which are
Bailey {\it ab}-type RR Lyrae's (RRab's) and 7 are {\it c}-type
(RRc's). The period distribution of the presently confirmed sample of
RRab's peaks at P=0\fd60, with a minimum period of 0\fd54. The
pulsational properties indicate for \leoi~ an intermediate Oosterhoff
type, similar to other dwarf galaxies in the Local Group and the
LMC.  However, the rather long minimum period of the {\it ab}-type
variables, and the significant number of RRab's with long period and
large amplitude, suggest that the bulk of the old population in \leoi~
is more like the Oosterhoff type II globular clusters. 
The most straightforward interpretation is that a range in metallicity
is present among the RR Lyrae's of \leoi, with a significant population
of very metal-poor stars. Alternatively, these OoII variables could be
more evolved.
The average apparent magnitude of the RR Lyrae's across the full cycle
is $\langle V(RR)\rangle= 22.60 \pm 0.12$ mag, yielding a distance
modulus $(m-M)_{V,0}= 22.04\pm 0.14$ mag for \leoi~ on the ``long''
distance scale.
\end{abstract}

%

\section{introduction}
\label{s_intro}

RR Lyrae have long been recognized to be excellent tracers of old
stellar populations, as well as good distance indicators for
Population II systems. 
%
RR Lyrae variables and/or extended horizontal branches (HB's) are now
known to exist in all Local Group dwarf spheroidal (dSph) galaxies,
despite the different and complex star formation histories
(Mateo \cite{mate98}; Da Costa \cite{daco98}).
The \leoi\ 
dSph was believed to represent the only exception to this
general trend,
with its 
predominantly ``young'' population, and a wealth of intermediate age
He-burning stars giving rise to a conspicuous red clump 
(Lee et al. \cite{mglee+93}; Caputo et
al. \cite{capu+98}; Gallart et al. \cite{gall+99a}, \cite{gall+99b}).
Recently, color-magnitude diagrams (\cmd's) extended to the outer
regions have revealed a horizontal branch well populated from blue to
red (Held et al. \cite{held+00}).
%
%
%
Given the presence of an extended HB and the low metallicity of the
system ([Fe/H]$\sim -2$, Lee et al. \cite{mglee+93})
one would naturally expect to find RR Lyrae variables in
\leoi. Indeed, a few stars near the limit of photographic
photometry were mentioned by Hodge \& Wright
(\cite{hodg+wrig78}) as probable RR Lyrae's caught at maximum light.
Unpublished work of Keane et al.  (\cite{kean+93}) also reports the
presence of these variables.

To confirm these early suggestions, a search for RR Lyrae and other
short period variables in \leoi\ was undertaken using the CCD mosaic
imager at the 2.2m ESO-MPI telescope. In this letter we report on the
successful detection of a conspicuous number ($\gtrsim$ 70) of RR
Lyrae variables in this galaxy.
%
The identification of RR Lyrae's in \leoi\ and the determination of
their average luminosity and pulsational properties provide us with
important tools for studying the metallicity and age of the oldest
stellar population and the early star formation history of \leoi, by
allowing at the same time a new independent measurement of the
distance to this galaxy.

\section{observations and reduction}
\label{s_obse} 
Images of \leoi~ ($\alpha _{2000}$=10 08 27, $\delta _{2000}$=12 18
30) through the $B$, $V$, $I$ filters were obtained on the nights of
April 20-25, 2000 with the Wide Field Imager (WFI) at the 2.2 m
ESO-MPI telescope at La Silla, Chile.
The WFI camera consists of 8 ``{\it ccd~44}''-type EEV CCD's with a total
field of view $34\pp \times 33\pp $. This coverage allowed us to
observe the body of the galaxy and its outer regions in just one
pointing.
%
The main body of the galaxy was centered on the lower four CCD's of
the mosaic to reduce light contamination from
Regulus, which is about 20 arcmin south.  The full data-set comprises
22 $B$, 40 $V$, and 4 $I$ 15-min exposures totaling $\sim20$ hours of
observing time spread over 6 consecutive half nights. 
Weather conditions were
generally photometric, while the seeing varied considerably during the
observations.
Standard stars from Landolt (\cite{land92}) were observed during
all photometric nights of the run in all CCD's.

Basic reduction (flat fielding, registration, co-adding) of the
CCD mosaic data were performed using the IRAF 
\footnote{ 
IRAF is distributed by the National Optical Astronomy Observatories,
which are operated by the Association of Universities for Research in
Astronomy, Inc., under cooperative agreement with the National Science
Foundation.}
package \mscred\ (Valdes \cite{vald98}) and the pipeline script
package \wfpred\ 
developed at the Padua Observatory. 
Reductions were done, and data calibrated, for the individual CCD's,
yielding calibration uncertainties of 0.03 mag in $B$ and 0.04 mag in
$V$ (night-to-night rms scatter of the zero points).  For variable
stars, the calibration were applied only to average magnitudes and
colors given the relatively large color terms of the WFI camera (about
0.28 and $-$0.075 in $b$ and $v$, respectively).
The \allframe\ program (Stetson \cite{stet94}) was used to measure all
images and combine the individual photometric measurements into a
master catalog for each CCD.  The combined photometry reaches
$V\simeq25.5$ (about 3 mag fainter than the HB) with errors $\sigma_V
= \sigma_B = 0.2$ mag (standard error of the mean).  The limiting
magnitude for the individual frames varies between $V=23.0$ and 24.5
depending on the seeing and photometric conditions. Measurement errors
vary accordingly between $\sigma_V=0.12$\,--\,0.35 at $V\sim23.5$.
The measurement errors for stars at the HB level, estimated from the
standard deviations of the measurements for non-variable stars, are
about $\sigma_V{\rm (HB)}=0.13$ and $\sigma_B{\rm (HB)}=0.11$ on the
individual frames.

\section{RR Lyrae variable stars in Leo\,{\sc i}}
\label{s_varsearch}

The time-series photometric measurements in the CCD catalogs were
analyzed to detect variable objects and derive their light curves.
Here we report on the first successful results for two CCD's (named
\#6 and \#7 in the ESO WFI Manual) comprising the body of the galaxy
(\leoi\ was centered on CCD \#7). All results for the remaining CCD's,
together with photometric measurements and light curves, will be
presented in a separate paper (G.~Clementini et al. \citedraft{2001}, in
preparation).
Identification of the variables was performed independently on the
instrumental $v$ and $b$ data sets, running the program \varfind\
(developed by Dr. P. Montegriffo) on the photometric catalogs (see
Clementini et al. \cite{cle+01a} for details).
%
%
At this stage, the search was optimized to find candidate variables
near the HB. Stars whose standard deviations of the $v$ and $b$
measurements are larger than 3$\sigma$, where $\sigma$ is the rms of
non-variable stars at the HB level, were flagged as candidate variables.
As a result, 96 candidates were found, some of which are in
the innermost area of \leoi.  
{\it RR~Lyrae's allow us to probe the old population all the way to
the center of the galaxy}.

%
Period search and definition of the light curve of the suspected
variables were done using differential photometry in the {\it
instrumental} magnitude system, with respect to some carefully chosen,
photometrically stable reference star (in general, red giants with
$B-V \geq 0.6$).
%
%
Period search was performed using \gratis\ (GRaphical
Analyzer of TIme Series), a software developed at the Bologna
Observatory (see Clementini et al. \cite{cle+00}, \cite{cle+01a}).
%

\realfigure{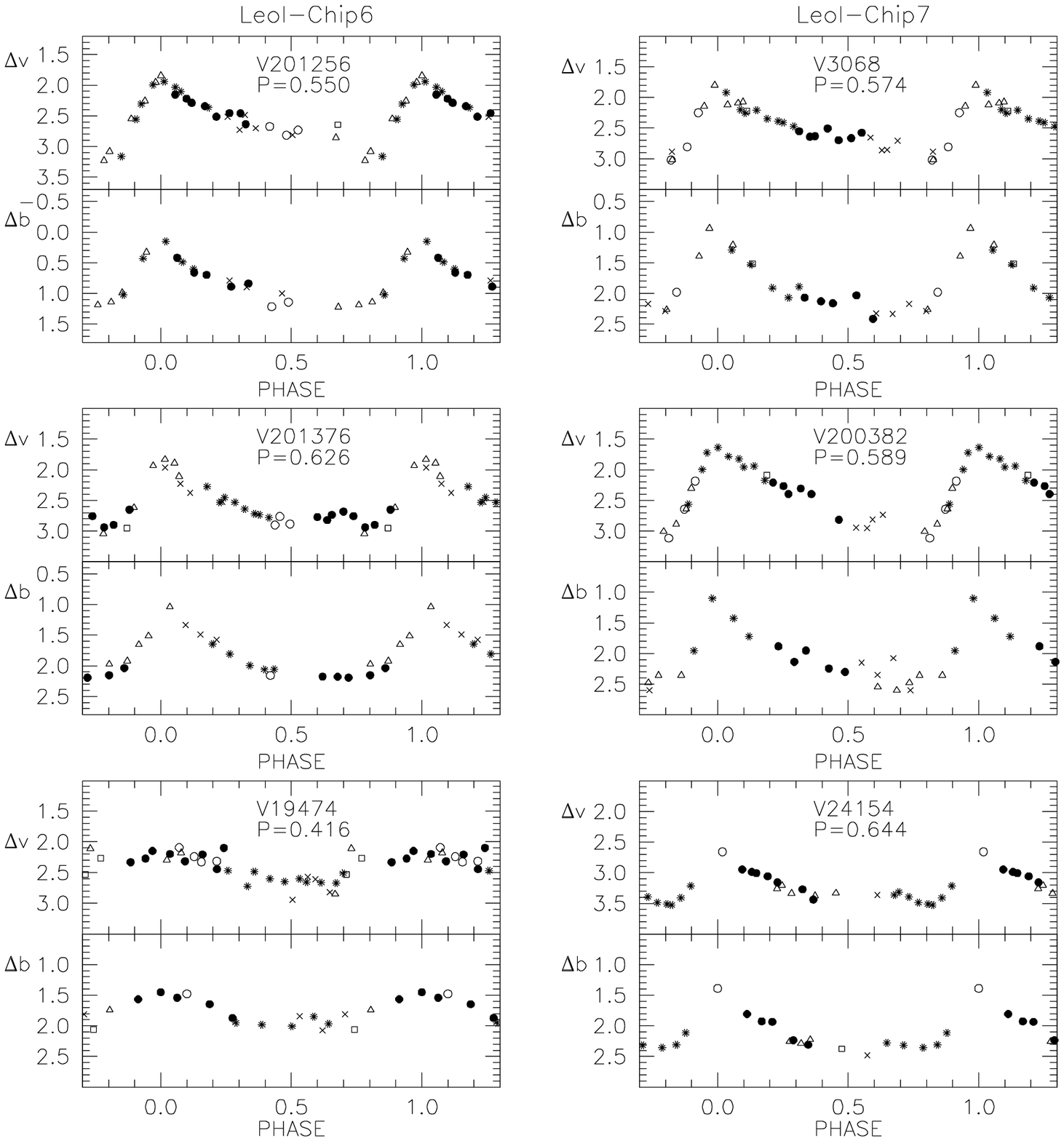}{
Differential $v$ and $b$ light curves of {\it ab}- and {\it c}-type RR
Lyrae's in \leoi~ from data in CCD \#6 (left panel) and \#7 (right
panel). Different symbols are used for data taken in different
nights. \label{f_lightc}}

Figure~\ref{f_lightc} presents the first light curves of RR Lyrae
stars in \leoi. We have plotted the differential $v$ and $b$ light
variations for a selection of RR Lyrae stars that sample quite well
the typical periods.
Out of 96 candidate variables, the majority were found to be {\it
ab}-type RR Lyrae's (63 stars), with a few examples of {\it c}-type RR
Lyrae's (11 objects).  We also identified one probable anomalous
Cepheid and 6 candidate short period binaries, while the remaining 15
objects could not be classified unambiguously.  Some
low amplitude variables, in particular {\it c}-type RR Lyrae's, may
actually have escaped detection.  A deeper search is in progress using
the Alard's (\cite{al00}) Optimal Image Subtraction method (G.~Clementini
et al. \citedraft{2001}, in prep.).
Periods, amplitudes, and epochs of maximum light were derived for a
subset of 54 RR Lyrae's (47 RRab's and 7 RRc's) with complete light
curves, using the \gratis\ $\chi^2$ Fourier fitting routine (one
harmonic was used for the {\it c}-type RR Lyrae's and two to five
harmonics for the RRab's). 
%
The accuracy of the derived periods depends on the sampling of the
light curves and is in general of the order 0.001 d. The amplitude
uncertainties, estimated from the rms of the fit residuals, are
$\sim$0.10 mag. 
Intensity-average differential magnitudes were obtained by integration
over the entire pulsation cycle using the best fitting model curves.
By adding the instrumental magnitudes of the reference stars, we
obtained the mean $b$, $v$ magnitudes of the RR Lyrae's, from which
the mean $B$, $V$ magnitudes were calculated via the calibration
equations.
%

\realfigure{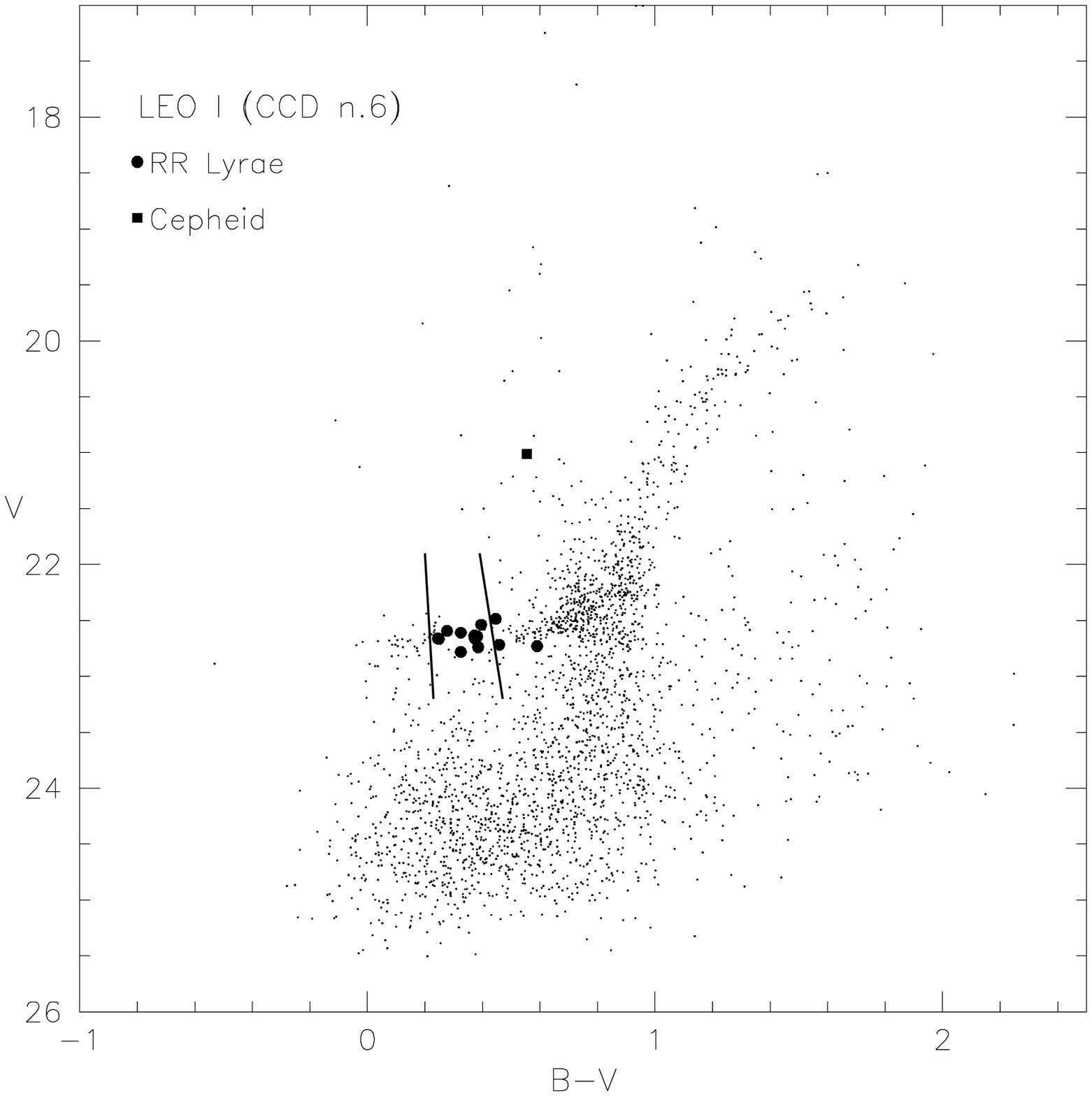} {
HR diagram of \leoi\ from a master \allframe\ combination of all 
data for CCD \#6. This \cmd\ is based on the
mean calibrated $B$, $V$ magnitudes of stars detected in both
passbands. The RR Lyrae variables (filled circles) and a Cepheid
(filled square) identified in this chip are plotted according to their
intensity-averaged magnitudes and colors. Lines show the blue and red
edges of the instability strip of the globular cluster M\,3, from
Corwin \& Carney's (\cite{cc01}) photometry, reddened to the value
appropriate for \leoi.  The red edge has been corrected to account for
the difference in metallicity with respect to \leoi\ (Walker
\cite{w98}).
\label{f_cmdchip6}}
%

The location of the detected variables in the HR diagram 
of the outer region of \leoi\ is shown in Figure~\ref{f_cmdchip6}. 
%
%
The vast majority of the data points for the RR Lyrae's are found on
the HB in correspondence to the instability strip, and their mean
magnitude defines the mean $V$ apparent luminosity of the HB
of \leoi.  



\section{Pulsational properties: 
clues to the old stellar population in Leo\,{\sc i}}
\label{s_pulprop}

Figure~\ref{f_pav} shows the relation between the period and the $V$
amplitude for the RR Lyrae's in the present sample.  Open symbols
represent the {\it c}-type pulsators (RRc), filled symbols are RRab
stars.  The average period of the \leoi~ RRab variables is $\langle
P_{ab}\rangle$=0\fd602 ($\sigma$=0\fd059), and the minimum
period is 0\fd539.  The solid lines in Figure~\ref{f_pav} represent
the $A_V - \log P$ relations defined by the {\it ab}-type variables
with clean light curves in the globular clusters M\,3 and M\,15
([Fe/H]=$-$1.66 and [Fe/H]=$-$2.15, respectively, on the Zinn \&
West's \cite{zw84} scale). These clusters can be considered the
prototypes of Oosterhoff type I (OoI) and II (OoII) globular clusters 
(GCs) in the
Milky Way (Sandage, Katem, \& Sandage \cite{sks81}; Sandage
\cite{s93}).  Also shown are the
$A_V - \log P$ relations followed by the {\it ab} variables in M2,
$\omega$~Cen, and NGC\,6441 ([Fe/H]=$-$1.62, $-$1.60, and $-$0.5 dex).
The data are from Clement \& Shelton (\cite{cs99}) for M3, Bingham et
al. (\cite{bingh+84}) and Silbermann \& Smith (\cite{ss95}) for M15,
Lee \& Carney (\cite{lc1+99a}) for M\,2, Clement \& Shelton
(\cite{cs99}) for $\omega$~Cen, and Pritzl et al. (\cite{pri+00}) for
NGC\,6441.

\realfigure{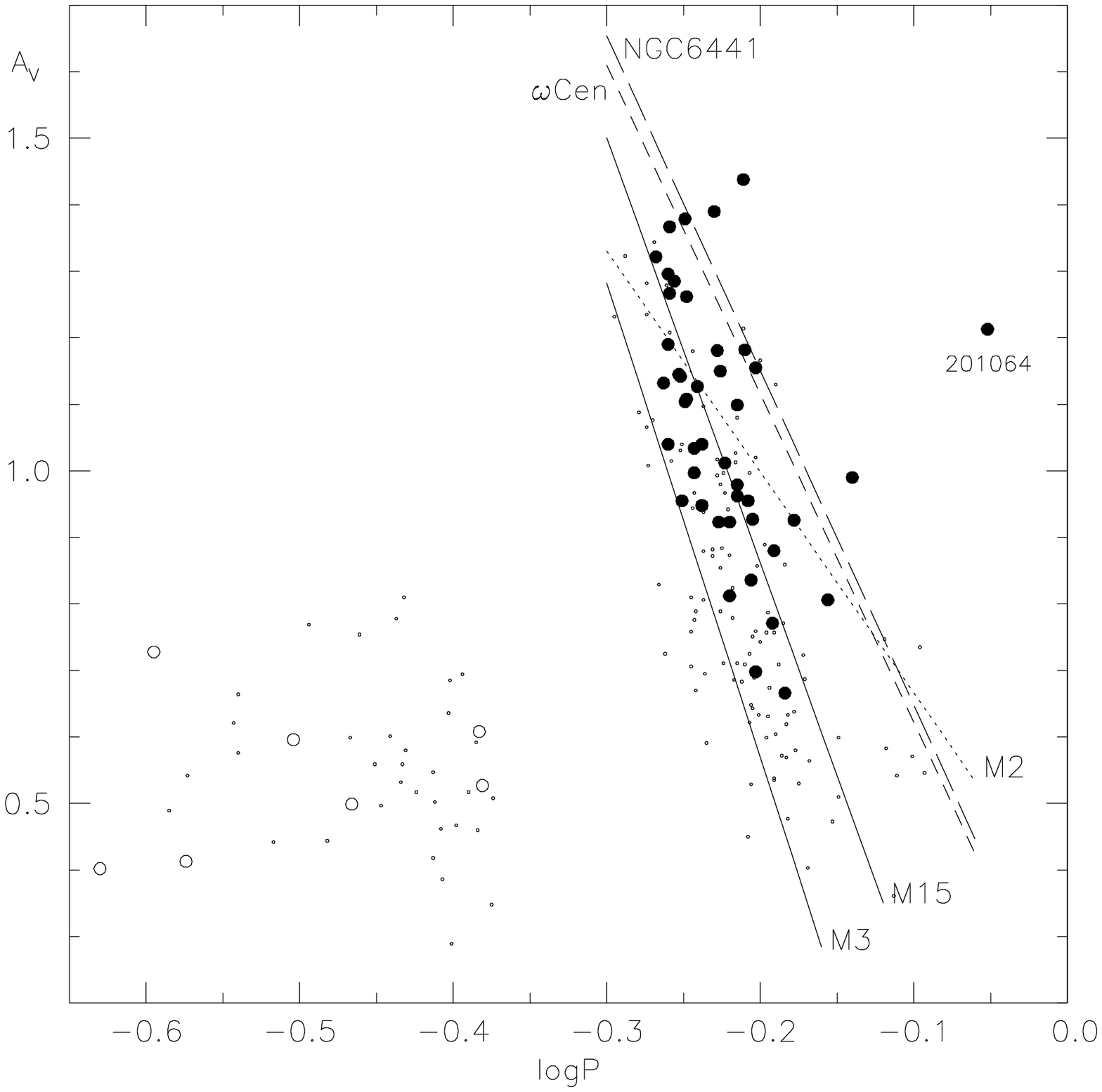} {Period--$V$ amplitude relation for the
RR Lyrae's in \leoi.  Periods are in days.  Filled and open symbols
are RRab and RRc variables, respectively.  The regression lines show
the $A_V - \log P$ relations for the {\it ab}-type variables 
in the globular clusters M\,3 and M\,15 (solid
lines), M\,2 (dotted line), $\omega$ Cen (dashed line), and NGC\,6441
(long dashed line).  Dots represent the amplitude-period
distribution of RR Lyrae variables in \leoii\ (Siegel \& Majewski
\cite{sieg+maje00}). \label{f_pav}}

Figure~\ref{f_pav} also shows a comparison with the 
$A_V - \log P$ distribution of
the RR Lyrae variables in the dSph galaxy \leoii\ (Siegel \& Majewski
\cite{sieg+maje00}). The pulsational properties of the RR Lyrae stars
in the two dwarf spheroidals appear quite similar.
The characteristics of the {\it ab}-type RR Lyrae's qualify \leoi~ as
a system intermediate between the OoI and OoII clusters, similar in
this respect to other dwarf galaxies in the Local Group (Siegel \&
Majewski \cite{sieg+maje00}; Cseresnjes \cite{cser01}; and references
therein) and to the LMC (Bono, Caputo, \& Stellingwerf \cite{bcs94}).
However, both the rather long minimum period of RRab's and the
presence of RRab's with long periods -- large amplitudes, indicate a
larger similarity with the OoII GCs.

The implications of such Oosterhoff intermediate properties 
for the old stellar populations in \leoi\ (and in
other dSph's as well) rests on the usual interpretation of
metallicity as the dominant parameter in determining 
the Oosterhoff dichotomy.
Indeed, OoII clusters are generally very metal-poor while OoI clusters
are of intermediate metallicity (e.g., Smith \cite{sm95}). 
If this interpretation is correct, then the distribution of
periods and amplitudes of \leoi\ RR Lyrae's 
implies a metallicity distribution for the old population 
extending from values more metal-poor than [Fe/H]=$-$2.15 dex (the
metallicity of the OoII cluster M\,15), to as metal-rich as
[Fe/H]$\sim-1.6$ (the metallicity of the OoI cluster M\,3).

Alternative interpretations may exist, though.
There are clusters like M\,2, $\omega$ Cen,
NGC\,6441, and NGC\,6388 whose positions in the period-amplitude
diagram imply that their RR Lyrae's are brighter than expected for
their metallicity (see Figure~\ref{f_pav}).  
 This observational evidence is interpreted as the Oosterhoff
 dichotomy being caused by evolution off the zero-age horizontal
 branch (ZAHB), with the OoI variables being on the ZAHB and the OoII
 variables being more evolved (Clement \& Shelton \cite{cs99}, Clement
 \& Rowe \cite{cr00}, Lee \& Carney \cite{lc2+99b}). Then, a
 difference in age would exist between the two Oosterhoff groups.
 Whether this can be a general interpretation of the Oosterhoff
 phenomenon or these are just peculiar clusters is still a matter of
 debate.
%
%
In this regard, it is interesting to note the presence in \leoi\ of a
variable with very long period (star 201064 in Figure~\ref{f_pav},
$P=0\fd889$). Although its large amplitude may be in doubt since the
variable falls near the center of \leoi\ where crowding is severe, and
is possibly affected from straylight from Regulus, 
variables with long periods have been found in NGC\,6441 and
NGC\,6388 (Pritzl et al. \cite{pri+00}).
%

A good indication on the mean metallicity of RR Lyrae's in \leoi\ can
be derived using the relation derived by Sandage (\cite{s93}) between
the average period of the RRab variables in the Milky Way GCs
and their metallicity, 
$\log \langle P_{ab}\rangle = -0.092 \mbox{[Fe/H]} - 0.389$.  
By applying this relation to the mean period of RRab
variables in \leoi, we obtain [Fe/H]$_{\langle P_{ab}\rangle}=-1.82$
for the average metallicity of the old population.
%
Previous values of the metallicity of \leoi\ are quite uncertain,
ranging from [Fe/H]=$-2$ to $-1$ (see Lee et al. \cite{mglee+93} and
references therein).  A recent comparison of the $(V-I)$ color of the
red giant branch (RGB) of \leoi\ with the fiducial sequences of
Galactic GCs yields a mean metallicity 
[Fe/H]$\simeq -1.8$ (Y.~Momany et al. \citedraft{2001}, in preparation).
%
Since the bulk of the red giant stars in \leoi\ have intermediate age
(between 2 and 7 Gyr: Caputo et al. \cite{capu+98}; Gallart et
al. \cite{gall+99b}), this value should be corrected (made more
metal-rich) by $\sim0.2$ dex 
(cf. Held, Saviane, \& Momany \cite{held+99}).
The fact that the RR Lyrae's are only slightly more metal-poor than
RGB stars seems to imply that metal enrichment (from the primordial
gas composition to [Fe/H]$\sim-1.8$) occurred early in the life of the
galaxy, and modest enrichment took place in between the early burst
that gave rise to the oldest population, and the main star formation
episode occurred $4 \pm 3$ Gyr ago.

\section{The distance to Leo\,{\sc i}}
\label{s_distance}

The mean magnitude of the RR Lyrae's in \leoi\ provides an independent
method to estimate the distance to this galaxy with some degree of
confidence.  The average apparent luminosity of the RR Lyrae's with
full coverage of the $B$ and $V$ light curves is $\langle V(RR)\rangle=
22.60 \pm 0.12$ mag (standard deviation of the mean, 48 stars). This
dispersion is fully accounted for by the combined effects of
photometric errors, metallicity distribution in the RR Lyrae's sample,
and evolution off the ZAHB.  Given the
large distance of \leoi, the spread caused by the intrinsic depth of
the galaxy is insignificant.

The absolute magnitude of the RR Lyrae's is $M_V{\rm (RR)}\simeq 0.50$
at [Fe/H]$=-1.5$ in the ``long'' distance scale, consistent with
$(m-M)^{\rm RR}_0$ = 18.53 for the LMC (see Walker \cite{pg99}, and
references therein), and 0.75 mag in the ``short'' scale. To correct
the absolute magnitude of the RR Lyrae's to the metallicity of the
\leoi\ variables, we adopted the RR Lyrae luminosity--metallicity
dependence given by the relation $\Delta M_V{\rm (RR)}/\Delta{\rm
[Fe/H]} = 0.2$ mag/dex, which is supported by both the Baade-Wesselink
analysis of Fernley et al. (\cite{fer+98}) and recent results on the
M\,31 globular clusters (Corsi et al. \cite{cor+00}).  For the
metallicity of \leoi\ we assumed the value derived in this paper from
the mean period of $ab$-type RR Lyrae's, [Fe/H]=$-1.82$.  
A reddening $E(B-V)=0.04 \pm 0.02$ mag was adopted
following Schlegel, Finkbeiner, \& Davis (\cite{schl+98}).  These
assumptions yield a distance modulus
$(m-M)^{\rm RR}_0 = 22.04 \pm 0.14$ mag ($d$=256 kpc) 
in the long scale (and 21.79 mag or 228 kpc in the short scale).  The
quoted error includes the errors on the mean RR Lyrae magnitude, the
calibration zero point error, and the adopted reddening uncertainty.
This result is in good agreement with our recent distance
determination from the RGB tip method on a different data set (Y.~Momany
et al. \citedraft{2001}, in prep.), yielding $(m-M)^{\rm tip}_0 =
21.93 \pm 0.16$.
%
The RGB tip method is based on the RR Lyrae
luminosity-metallicity relation of Lee, Demarque, \& Zinn
(\cite{lee+90}), yielding $M_V{\rm (RR)} = 0.565$ at
[Fe/H]$=-1.5$. 
Our new result is intermediate between previous values
in the literature (see Lee et al. \cite{mglee+93}).


\acknowledgments 

We thank P. Stetson for providing us his set of photometric
programs, including \allframe.  We are grateful to M. Catelan for
helpful discussions regarding the nature of the Oosterhoff dichotomy,
to C. Cacciari for useful comments on the manuscript, and 
to an anonymous referee for helpful remarks. This study
was partially supported by the National projects ``Processing of
large-format astronomical data'' (COFIN-98) and ``Stellar Observables
of Cosmological relevance'' (COFIN-00).


\end{document}